\documentclass[twocolumn,showpacs]{revtex4}
\usepackage{epsfig}
\usepackage{amsmath}
\usepackage{amsfonts}
\usepackage{amssymb}
\usepackage{graphicx}

\def\nn{\nonumber}

\def\J#1#2#3#4{#2 {\bf #3} #4 (#1)}
\def\PTP{Prog. Theor. Phys.}
\def\PRL{Phys. Rev. Lett.}
\def\PRD{{Phys. Rev.} D}
\def\PRE{{Phys. Rev.} E}
\def\PR{Phys. Rev.}

\def\APL{Ann. Phys. (Leipzig)}

\def\JMP{J. Math. Phys.}
\def\CQG{Class. Quantum Grav.}

\def\MNRAS{Mon. Not. R. Soc.}

\begin{document}
\title{{\bf Chaotic dynamics around
astrophysical objects with nonisotropic stresses}}

\author{F. L. Dubeibe}
\email{fldubeibem@unal.edu.co} \affiliation{Escuela de F\'isica,
Universidad Industrial de Santander, A.A. 678, Bucaramanga,
Colombia}\affiliation{Departamento de F\'isica, Universidad Nacional
de Colombia, Santa f\'e de Bogot\'a D.C., Colombia}

\author{Leonardo A. Pach\'{o}n}
\email{lapachonc@unal.edu.co} \affiliation{Escuela de F\'isica,
Universidad Industrial de Santander, A.A. 678, Bucaramanga,
Colombia} \affiliation{Departamento de F\'isica, Universidad
Nacional de Colombia, Santa f\'e de Bogot\'a D.C., Colombia}

\author{Jos\'e D. Sanabria-G\'omez}
\email{jsanabri@uis.edu.co} \affiliation{Escuela de F\'isica,
Universidad Industrial de Santander, A.A. 678, Bucaramanga,
Colombia}

\date{\today}

\begin{abstract}
The existence of chaotic behavior for the geodesics of the test
particles orbiting compact objects is a subject of much current
research. Some years ago, Gu\'eron and Letelier [Phys. Rev. E
\textbf{66}, 046611 (2002)] reported the existence of chaotic
behavior for the geodesics of the test particles orbiting compact
objects like black holes induced by specific values of the
quadrupolar deformation of the source using as models the
Erez--Rosen solution and the Kerr black hole deformed by an internal
multipole term. In this work, we are interesting in the study of the
dynamic behavior of geodesics around astrophysical objects with
intrinsic quadrupolar deformation or nonisotropic stresses, which
induces nonvanishing quadrupolar deformation for the nonrotating
limit. For our purpose, we use the Tomimatsu-Sato spacetime [Phys.
Rev. Lett. \textbf{29} 1344 (1972)] and its arbitrary deformed
generalization obtained as the particular vacuum case of the five
parametric solution of Manko et al [Phys. Rev. D 62, 044048 (2000)],
characterizing the geodesic dynamics throughout the Poincar\'e
sections method. In contrast to the results by Gu\'eron and Letelier
we find chaotic motion for oblate deformations instead of prolate
deformations. It opens the possibility that the particles forming
the accretion disk around a large variety of different astrophysical
bodies (nonprolate, e.g., neutron stars) could exhibit chaotic
dynamics. We also conjecture that the existence of an arbitrary
deformation parameter is necessary for the existence of chaotic
dynamics.

\end{abstract}

\maketitle
\section{Introduction}
\label{Introduction}

Although it is very usual in the literature that the stress tensor
of several astrophysics objects matter --neutron stars, exotic
stars-- is approximated by that of a perfect fluid, {\it viz.}
nonviscous medium of total energy density ${\cal E}$ (mass density
$\rho = {\cal E}/c^2$), in which all stresses are zero except for
an isotropic pressure $P$. We have to notice that in astrophysics,
the most common object are that which present nonvanishing
quadrupolar deformation for the nonrotating limit, i.e., objects
not deformed by current mass --rotation-- either by arbitrary
multipoles --arbitrary mass quadrupole or mass octupole-- but
intrinsically deformed. Besides, it is clear that the
approximations is based on the fact that the shear stresses, e.g.,
those produced by elastic strain in the solid crust or by strong
magnetic field, are generally negligible compared to the pressure,
but it is also clear that when the pressure is assumed isotropic
we refuse the possibility of intrinsic deformation because it
could obey to nonisotropic stresses during its formation process.

With the aim to follow the usual premise that the exterior
gravitational field of the relevant astrophysical objects can be
modelled by stationary axially symmetric exact solutions to the
Einstein--Maxwell field equations (see e.g. \cite{PRDAMNS, Mielke}
for the case of neutron stars) we adopt spacetimes endowed naturally
in the Weyl--Lewis--Papapetrou line element \cite{Papap}. This
metric admits two Killing vectors, one timelike and other spacelike,
which correspond to the desiderated symmetries mentioned above. In
the literature, we can find several exact solutions to the
stationary axis--symmetric Einstein--Maxwell system equations (see
\cite{Kramer} for some of them), but just a few of them described
objects intrinsically deformed. The most known are the corresponding
members to the Tomimatsu--Sato family \cite{TSPRL}, \cite{SlowlyJD}
which quadrupole deformation in the nonrotating limit is ${\cal
Q}=-\frac{2}{\delta^3}m^3$, being $m=\frac{\delta M_0 }{2}$ where
$M_0$ is the mass monopole of the source and $\delta$ a
dimensionless parameter, taking the values $\delta = 1$ for the Kerr
solution and $\delta = 2$ for the Tomimatsu--Sato $\delta = 2$
solution.

In order to clarify the physical interpretation of the
Tomimatsu--Sato family (henceforth {\small \sc {TS}}-family), we
present briefly in the next paragraphs some interpretations given to
it.

In the early 1970's, when the solution was obtained, Tomimatsu \&
Sato \cite{TSPTP} and after Tanabe \cite{Yukio} showed that {\small
\sc {TS}}-family represents the gravitational field of rotating
masses with angular momentum about the $z$-axis. But, soon was
realized that {\small \sc {TS}}-family suffers of naked curvatures
singularity. Papadopoulos and Xanthopoulos \cite{Papadopulus} tried
to resurrect interest in these solutions putting them in other
context, they modified slightly the {\small \sc {TS}}-metrics by a
suitable analytic continuation and concluded that these solutions
represent cylindrical symmetric spacetime, interpreting this time
the solution as a beamlike-shaped pulse of gravitational radiation
scattered by a cosmic string. Kodama \& Hikida \cite{Kodama} showed
that the two points in the Weyl coordinates, which have been
recognized as the directional singularities, are really
two-dimensional surfaces and that these surfaces are horizons. They
also showed that each of the two horizons has the topology of a
sphere and concluded that this may indicates that {\small \sc
{TS}}-family describes the spacetime surrounding a new possibility
of final states of gravitational collapse.

In 2002, using the Sibgatullin's integral  method
\cite{SibgatullinMethod}, Mielke, Manko and Sanabria--G\'omez
generalized a member of the vacuum {\small \sc{TS}}-family, the
Tomimatsu--Sato $\delta=2$ solution (henceforth {\small \sc{TS2}}),
to a most general spacetime in the electrovacuum case
\cite{SlowlyJD} possessing parameters for the electric charge $Q$,
magnetic dipole $\mu$, and arbitrary quadrupolar deformation $b$.
Besides, Berti \& Stergioulas \cite{Berti} and Berti, White,
Maniopoulou \& Bruni \cite{BertiBruni} studied this solution as
possible model to describe the gravitational field of a Rapidly
Rotating Neutron Stars (henceforth {\small \sc {RRNS}}). In that
work the generalized version of the {\small \sc {TS2}} solution
\cite{SlowlyJD} was considered and it was concluded that this
solution is a suitable model for the field of a neutron star but
only in the faster rotation regime. The reason why the solution by
Manko {\it et al.} cannot be used to model neutron stars close to
the nonrotating limit, (i.e.,  when the source is slightly deformed
--quasispherical) can be understood if we take into account that in
this limit, neglecting the high order multipoles, \cite{SlowlyJD}
reduces to {\small \sc{TS2}} leaving a nonvanishing deformation
given by ${\cal Q}=-\frac{1}{4}m^3$, which differs substantially of
the spherical symmetry due to the high mass of the star.

In this point, following the concluding remarks by Kodama {\it et
al.} \cite{Kodama} and Berti {\it et al.} \cite{Berti}, we assume
that {\small \sc{TS}}--family can be used to describe the topology
of the spacetime around relevant astrophysical objects like
neutron stars, strange quarks or any other exotic final states of
the gravitational collapse. This last statement and the fact that
the members of the {\small \sc{TS}}--family have non--vanishing
quadupolar deformation constitute the initial point for our work
because it enables us to consider the members of the {\small
\sc{TS}}--family as analytic closed form solutions for the
gravitational field of relevant astrophysical objects with
anisotropic stresses tensor.

After having an analytic closed form solution for the gravitational
field of a source, one of the interesting topic to study is the
motion of the particles orbiting this source. The construction of
astrophysical models which are able to give us a complete
description of realistic astronomical systems has to take into
account the behavior of the surrounding matter in order to compare
with astronomical observation and emit a proposition about the
validity of the model. In this work, we are interested in the nature
of the dynamics --chaotic or regular-- of the test particles which
orbit two specific members of the {\small \sc{TS}}--family, the
{\small \sc{TS2}} and the solution by Manko {\it et al.} in the
vacuum case, {\it viz}. $Q=\mu=0$.

In general relativity the study of stochastic motions in
deterministic system --deterministic chaos-- has followed two main
branches. The first one is the study of the geodesic motion of test
particles in a given gravitational field (Bombelli and Calzetta
\cite{Bombelli}, Vieira and Letelier \cite{LetelPRL}\cite{LetelPRD},
Gu\'eron and Letelier \cite{LetelPRE} and references therein). The
other branch is the time evolution of the gravitational field itself
(Motter and Letelier \cite{MotterPRD}, Hobill, Burd and Coley
\cite{Hobill}), which is relevant in cosmology (see e.g. Motter
\cite{MotterPRL}). This work is in line with the first scheme and is
organized as follows. In Sec. II the {\small \sc {TS2}} solution and
the Manko {\it et al.}' solution are presented and a brief
discussion of their features is given. In Sec. III the dynamics of
geodesics of test particles is analyzed. Finally, in Sec. IV a brief
discussion about the obtained results is presented.

\section{The particular space-times}
\label{analgravfield}

\subsection{Case I: The Tomimatsu--Sato $\delta=2$ solution}
The most simple form of the metric for a stationary axisymmetric
space time was given by Papapetrou \cite{Papap} and it can be
written as
\begin{equation}
    \label{Papapetrou}
    ds^2= f(dt-\omega d\varphi)^2 - f^{-1} [e^{2\gamma}(d\rho^2 +
    dz^2) + \rho^2 d\varphi^2],
\end{equation}
here $f$, $\omega$ and $\gamma\,$ are functions of the
quasicylindrical Weyl-Lewis-Papapetrou coordinates $(\rho,z)$. The
Weyl-Lewis-Papapetrou coordinates are related with the prolate
spheroidal coordinates $(x,y)$ by means of the transformation
\begin{equation}
    \label{2}
    \rho^{2}=k^{2}(x^{2}-1)(1-y^{2})\, ,\qquad z=k x y\, ,
\end{equation}
with $ x\geq 1 \,\,{\rm{and}}\,\, -1\leq y \leq 1\, ,$ then the
metric is rewritten as
\begin{eqnarray}
\label{metric_spheroidal} ds^2 &=& f \, (dt - \omega d\varphi)^2 -
k^2 f^{-1} \Bigl[ e^{2 \gamma} \,
(x^2 - y^2) \Bigl( \frac{dx^2}{x^2 - 1} \nonumber \\
&& + \frac{dy^2}{1 - y^2} \Bigr) + (x^2 - 1) (1 - y^2) d\varphi^2
\Bigr].
\end{eqnarray}
The {\small \sc TS} metrics  were obtained from a solution to the
Ernst equation in the vacuum case \cite{ErnstV}, which is given by
\begin{equation}
    \label{xiTS}
    \xi = \frac{p^{2}x^{4}+q^{2}y^{4}-2  i p q x
    y(x^{2}-y^{2})-1}{2px(x^{2}-1)-2 i q y (1-y^{2})}\, ,
\end{equation}
where $p=(1-q^{2})^{1/2}$, $q=J/m^{2}$ and $k=m p/\delta$. The
$\delta$  parameter is dimensionless, taking the values $\delta =
1$ and $\delta = 2$ for the Kerr and the {\small \sc TS2}
solutions, respectively. The metric functions derived from
(\ref{xiTS}) are \cite{TSPRL}
\begin{equation}
     f=\frac{A}{B}\, ,\quad \omega= \frac{2 m q C (1-y^{2})}{A}\, ,\quad
     e^{2\gamma}=\frac{A}{p^{4}(x^{2}-y^{2})^{4}}\, ,
\end{equation}
where
\begin{eqnarray*}
A &=& p^{4}(x^{2}-1)^{4}+q^{4}(1-y^{2})^{4}
    -2p^{2}q^{2}(x^{2}-1)(1-y^{2})
\\
    &\times& \{2(x^{2}-1)^{2}+2(1-y^{2})^{2}+ 3(x^{2}-1)(1-y^{2})\}\, ,
\\
B&=&\{p^{2}(x^{2}+1)(x^{2}-1)-q^{2}(y^{2}+1)(1-y^{2})
\\
    &+&2p\,x(x^{2}-1)\}^{2}+4q^{2}y^{2}\{p\,x(x^{2}-1)
\\
    &+&(p\,x+1)(1-y^{2})\}^{2}\, ,
\\
C &=&-p^{3}x(x^{2}-1)\{2(x^{2}+1)(x^{2}-1)+(x^{2}+3)(1-y^{2})\}
\\
    &-&p^{2}(x^{2}-1)\{4x^{2}(x^{2}-1)
    +(3x^{2}+1)(1-y^{2})\}
\\
    &+&q^{2}(p\,x+1)(1-y^{2})^{3}\, .
\end{eqnarray*}
The physical sense of the parameters $m\, $ and $J$ is derived
form the Simon's multipole moments \cite{Simon}. The first three
of the relativistic multipole moments calculated from (\ref{xiTS})
with the aid of the Hoenselaers-Perj\'es procedure
\cite{HoensPerj} are
\begin{eqnarray}
    \label{mmasa}
    P_0 = \frac{2m}{\delta}, \quad
    P_1 = i\frac{4J}{\delta^2}, \quad
    P_2 = -\frac{2}{\delta^3}\left(1+\frac{3J^{2}}{m^{4}}\right)m^3.
\end{eqnarray}
The terms $P_0$ and $P_2$ denote  the monopole and the quadrupole of
mass, respectively. They are related to the mass and the deformation
of the source.  On the other hand, the term $P_1$ describes the
angular momentum.  From (\ref{mmasa}) is obvious that $P_0$, $P_2$,
and $P_1$ are determined only by two parameters, $m$ and $J$, and
means that the source only has mass and arbitrary angular momentum.

Defining the parameter $j$ as $J/m^2$ the quadrupole of mass for
the case $\delta=2$ can be written as $P_2={\cal Q}=-0.25\,
m^3(1+3j^2) \,.$ From the previous expression it is clear that the
quadrupole deformation of {\small \sc {TS2}} solution always is
negative. For such reason we can affirm that the {\small \sc
{TS2}} solution describes the spacetime around an oblate source.

\subsection{Case II: Solution by Manko, Mielke and Sanabria--G\'omez}

This solution has five relevant and independent parameters: $m$ the
gravitational mass, $a$ the specific angular momentum ($a=J/m$), $Q$
the electric charge, $b$ a parameter related with the mass
quadrupole moment, and $\mu$ a parameter related with the dipolar
magnetic moment. In addition, it has a remarkable feature, which is
that the quadrupolar deformation,
\begin{equation}
\label{deformation}
  {\cal Q}=m(\delta-d-a(a-b)),
\end{equation}
with
\begin{eqnarray}
\delta &=& \frac{\mu^2 - m^2 b^2}{m^2 - (a - b)^2 - Q^2}, \nonumber \\
d &=& \frac{1}{4} \, [ m^2 - (a - b)^2 - Q^2 ],
\end{eqnarray}
depends directly on electromagnetic parameter $Q$ and $\mu$. This
means that a test particle ``sees'' a source with a different
quadrupolar deformation than the real one. This feature could
implies that the electromagnetic field can induce chaos in the
dynamic geodesic for uncharged particles orbiting this source
\cite{note1}. In this work, we are interested in the influence of
the real deformation of the mass and not in the effective
deformation. For that reason, we will choose $Q=0$ and $\mu=0$,
then the metric functions take the form
\begin{eqnarray}
\label{metricfunctions_manko}
    f = \frac{E}{D}, \quad e^{2 \gamma} = \frac{E}{16 \kappa^8 (x^2 -
    y^2)^4}, \quad \omega = \frac{- (1 - y^2) F}{E}, \nonumber \\
\end{eqnarray}
with
\begin{eqnarray} \label{metricfunctions_abbr}
E &=& \{ 4 [ \kappa^2 (x^2 - 1) + \delta (1 - y^2) ]^2 + (a - b) \nonumber \\
&& [ (a - b) (d - \delta) - m^2 b ] (1 - y^2)^2 \}^2  \nonumber \\
&& - 16 \kappa^2 (x^2 - 1) (1 - y^2) \{ (a - b) [ \kappa^2 (x^2 -
y^2)
\nonumber \\
&& + 2 \delta y^2 ] + m^2 b y^2 \}^2, \nonumber \\
D &=& \{ 4 (\kappa^2 x^2 - \delta y^2)^2 + 2 \kappa m x [ 2 \kappa^2
(x^2 - 1) \nonumber \\
&& + (2 \delta + a b - b^2) (1 - y^2) ] + (a - b) [ (a - b) (d -
\delta)
\nonumber \\
&& - m^2 b ] (y^4 - 1) - 4 d^2 \}^2 + 4 y^2 \{ 2 \kappa^2 (x^2 - 1)
 \nonumber \\
&& [ \kappa x (a - b) - m b ] - 2 m b \delta (1 - y^2)+ [ (a - b) \nonumber \\
&& (d - \delta) - m^2 b ] (2 \kappa x + m) (1 - y^2) \}^2, \nonumber \\
F &=& 8 \kappa^2 (x^2 - 1) \{ (a - b) [ \kappa^2 (x^2 - y^2) + 2
\delta y^2 ] \nonumber \\
&& + m^2 b y^2   \} \{ \kappa m x [ (2 \kappa x + m)^2
- 2 y^2 (2 \delta + a b  \nonumber \\
&& - b^2) - a^2 + b^2] - 2 y^2 (4 \delta d
\nonumber \\
&& - m^2 b^2) \} + \{ 4 [ \kappa^2 (x^2 - 1) + \delta (1 - y^2) ]^2
\nonumber \\
&& + (a - b) [ (a - b) (d - \delta) - m^2 b ] (1 - y^2)^2 \} \nonumber \\
&& (4 (2 \kappa m b x + 2 m^2 b) [ \kappa^2 (x^2 - 1) + \delta (1 - y^2) ] \nonumber \\
&& + (1 - y^2)  \{ (a - b) (m^2 b^2- 4 \delta d) - (4 \kappa m x + 2 m^2   \nonumber \\
&& ) [ (a - b) (d - \delta) - m^2 b ] \} ) .
\end{eqnarray}

where $x = \frac{1}{2 \kappa} (r_{+} + r_{-})$ and  $y =
\frac{1}{2 \kappa} (r_{+} - r_{-})$ with $r_{\pm} = \sqrt{\rho^2 +
(z \pm \kappa)^2}$ and $\kappa = \sqrt{d+\delta}$. If, in
addition, we put $a=0$ and $b=0$ we have, obviously, that the
quadrupolar deformation (\ref{deformation}) corresponds to the
deformation for {\small \sc {TS2}} solution, so all the
consideration pointed out above are also valid for this case.

\section{Geodesics Dynamics for Test Particles}
\label{sec:potential}

Following an standard procedure, we define the Lagrangian function
${\cal L}$ as $2{\cal L}=g_{\mu\nu}dx^{\mu}/d\tau \,
dx^{\nu}/d\tau$, for the stationary axisymmetric metric
(\ref{metric_spheroidal}), we obtain
\begin{eqnarray}\label{L}
2{\cal L}&=&f(\dot{t}-\omega\dot{\varphi} )^2 - \frac{k^{2}}{f}
\Bigl[e^{2\gamma}(x^{2}-y^{2}) \Bigl (\frac{\dot{x}^2}{x^{2}-1} \nonumber \\
&& +  \frac{\dot{y}^2}{1-y^{2}}\Bigr) + (x^{2}-1)(1-y^{2})
\dot{\varphi}^2\Bigr]\, ,
\end{eqnarray}
where the overdot indicates  derivation respect the proper time
$\tau$. Using the Hamiltonian formalism we found that the motion
equations for the test particle are given by
\begin{eqnarray}
    \dot{x}&=&-\frac{f(x^{2}-1)}{k^{2}e^{2\gamma}(x^{2}-y^{2})}P_{x},
\quad
    \dot{y}=-\frac{f(1-y^{2})}{k^{2}e^{2\gamma}(x^{2}-y^{2})}P_{y},
\label{e1}
\nn\\
\\
    \dot{P_x} &=& -\frac{1}{2}\left\{\frac{E^2}{f}-
    \frac{f(L+E\omega)^{2}}{k^{2}(x^{2}-1)(1-y^{2})}\right.
\nn \\
    &-&\left.\frac{f}{k^{2}e^{2\gamma}(x^{2}-y^{2})} \left[P_{x}^2(x^{2}-1) +
    P_{y}^2(1-y^{2})\right]\right\}_{,\,x}
\label{e3}
\\
    \dot{P_y} &=& -\frac{1}{2}\left\{\frac{E^2}{f}-
    \frac{f(L+E\omega)^{2}}{k^{2}(x^{2}-1)(1-y^{2})}\right.
\nn\\
    &-&\left. \frac{f}{k^{2}e^{2\gamma}(x^{2}-y^{2})} \left[P_{x}^2(x^{2}-1) +
    P_{y}^2(1-y^{2})\right]\right\}_{,\,y}
\label{e4}
\end{eqnarray}
with
\begin{eqnarray}
    \label{3.5}
   E&=&f(\dot{t}-\omega\dot{\varphi})\, ,
\\
    L&=&-\omega f (\dot{t}-\omega\dot{\varphi})-
    \frac{k^{2}}{f}(x^{2}-1)(1-y^{2})\dot{\varphi}\, .
\end{eqnarray}
The constants of integration $E$ and $L$ are related to the energy
and to the angular momentum of the test particle, respectively. In
the case of timelike geodesic, the lagrangian ${\cal L}$ satisfies
${\cal L}=0.5$, this relation allows us to define an effective
potential, which explicitly is
\begin{equation*}\label{3.10}
\Phi(x,y)=\frac{f}{k^{2}e^{2\gamma}(x^{2}-y^{2})}\left[\frac{E^2}{f}
-\frac{f(L+E\omega)^{2}}{k^{2}(x^{2}-1)(1-y^{2})}-1\right].
\end{equation*}
By the $\Phi$ definition the motion must be restricted to the
region $\Phi\geq 0$. With the aim of study the dynamic of
geodesics is necessary to be sure that the test particles motion
is in a confinement region. The existence of such regions is
determined throughout the condition $\Phi\geq 0$.

The solution to the equation system (\ref{e1})-(\ref{e4}), could be
found using a symplectic Runge-Kutta method. That method let us find
the numerical solution of the system, given the constants $E,L$ and
the initial conditions $x(0),y(0),P_{x}(0)$ and $P_{y}(0)$. By the
existence of the integral of motion
${\mathcal{L}}={\mathcal{H}}=0.5$ with ${\mathcal{H}}$ the
hamiltonian of the system, if we have $E,L$ and
$x(0),y(0),P_{x}(0)$, the momentum $P_{y}(0)$ will be determined for
this equation. The values of $x(0)$ and $y(0)$ are selected in such
form that satisfies the condition $\Phi\geq 0$ for confined motions,
then the only arbitrary parameter is $P_{x}(0)$. These constants of
motion indicate to us that the geodesic motion is performed in a
three dimensional effective phase space in which the Poincar\'e
section method is an adequate tool to study the motion (see Letelier
\& Gueron \cite{LetelPRE,LetelAA}).

\subsection{Case I: {\small \sc \bf TS}2 solution}

\begin{figure}[!h]
\includegraphics[scale=0.63]{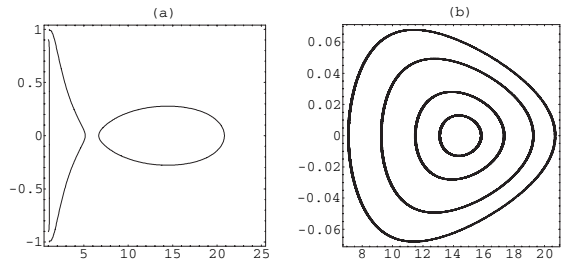}
\caption{(a) Boundary contour $\Phi=0$ using $E=0.94$, $L=-3.12$.
There is one escape zone in the left-hand side of the picture, which
correspond to small values of $x$, and a closed zone of bounded
motion to the right. (b) Poincar\'e sections in the plane $x\-P_{x}$
for the values defined in (a). We have regular motion.} \label{TS}
\end{figure}

First we shall to consider the {\small \sc TS}2 solution. In this
case, the $k$ parameter introduced in (\ref{metric_spheroidal}) is
completely determined by the mass and the angular momentum of the
source, implying that we only can vary the energy $E$, and the
angular momentum $L$, of the test particle. For this solution, we
only found bounded region of motion like that shown in
Fig.\ref{TS}(a). Any configuration with two or more bounded regions
of motion was not found. In Fig.\ref{TS}(a), the curve $\Phi=0$ in
the plane $x\-y$ for $E=0.94$, $L=-3.12$ is plotted and it is
observed that there just exits one bounded region and one escape
region.
In Fig.\ref{TS}(b) we can see that the motion in the bounded
region of the Fig.\ref{TS}(a) is completely regular such as that
which occurs in the Schwarzschild, Kerr and Kerr--Newman black
holes \cite{Carter}. The geodesics for this solution were studied
using surface sections for many different values of $E$ and $L$.
The numerical results suggest the existence of only integrable
geodesics.

\subsection{Case II: Manko {\it et al.} Solution}

In this case, we used the available information present in the
literature for typical values for the multipolar structure of
neutron stars. In particular, we took numerical data from the
Berti and Stergioulas work \cite{Berti}.  In \cite{Berti}, Berti
and Stergioulas  solved in a numerical way the full Einstein
equations to determine the spacetime for rapidly rotating neutron
star along of sequences of constant rest mass for selected
equation of state (henceforth EOS) denoted as EOS A \cite{EOSA},
EOS AU \cite{EOSAU}, EOS FPS \cite{EOSFPS}, EOS L \cite{EOSL} and
EOS APRb \cite{EOSAPRb}.
\begin{figure}[!h]
\includegraphics[scale=0.63]{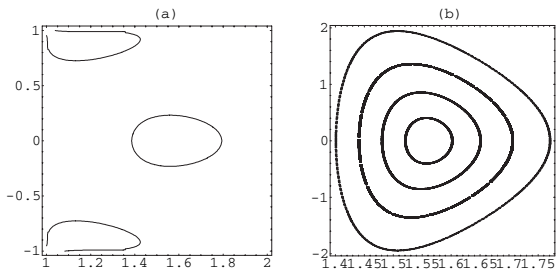}
\caption{(a) Boundary contour $\Phi=0$ for $M=1.840M_\odot$,
$J=3.683$, $b=-0.3792$, $E=0.1$ and $L=-5.6$. There is two escape
zones, and a closed zone of bounded motion. (b) Poincar\'e sections
in the plane $x\-P_{x}$ for the values defined in (a). We see only
regular motions.} \label{MS1}
\end{figure}
\begin{figure}[!h]
\includegraphics[scale=0.63]{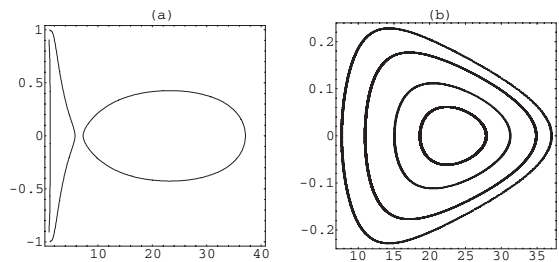}
\caption{(a) Boundary contour $\Phi=0$ for $M=1.936M_\odot$,
$J=4.498$, $b=-0.3080$, $E=0.95$ and $L=-9.0$. There is one escape
zone, and a closed zone of bounded motion. (b) Poincar\'e sections
in the plane $x\-P_{x}$ for the values defined in (a). The geodesics
are only regular.} \label{MS2}
\end{figure}
\begin{figure}[!h]
\includegraphics[scale=0.63]{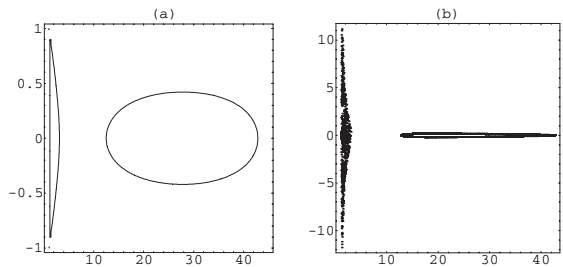}
\caption{(a) Boundary contour $\Phi=0$ for $M=1.936M_\odot$,
$J=4.498$, $b=0.8$, $E=0.96$ and $L=-9.9$. There are two small
escape zones, and two closed zones of bounded motion. (b) Poincar\'e
sections in the plane $x\-P_{x}$ for the values defined in (a). We
see chaotic motion to the left-hand side and regular motion to the
right-hand side.} \label{MS3}
\end{figure}
\begin{figure}[!h]
\includegraphics[scale=0.63]{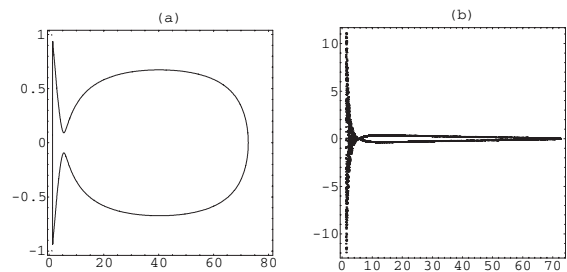}
\caption{(a) Boundary contour $\Phi=0$ for $M=1.936M_\odot$,
$J=4.498$, $b=0.8$, $E=0.971$ and $L=-9.3$. There are two small
escape zones, and two regions linked by a narrow connection of
bounded motion. (b) Poincar\'e sections in the plane $x\-P_{x}$ for
the values defined in (a). We see chaotic motion in the left-hand
side of the figure and in a small external region on the right-hand
side.} \label{MS4}
\end{figure}

Additionally, they matched the Manko {\it et. al.} solution
\cite{SlowlyJD} to the numerical solutions imposing the condition
that the mass--quadrupole moment of the numerical and analytic
spacetimes be the same. Under this condition, the $b$ parameter is
determined from the numerical data. From \cite{Berti} we can see
that the Manko {\it et al.} solution fixes better for the EOS FPS,
for that reason we took all the values presented by Berti and
Stergioulas for each sequence of mass of the EOS FPS founding
bounding potentials like the presented in Fig.\ref{MS1}(a) and
Fig.\ref{MS2}(a).

In Fig.\ref{MS1}(a), we present the boundary contour for $\Phi=0$
in the $x\-y$ plane for $M=1.840M_\odot$, $J=3.683$, $b=-0.3792$,
$E=0.1$ and $L=-5.6$, observing that exists two scape regions and
a bounded region. In Fig.\ref{MS2}(a) boundary contour for
$\Phi=0$ in the plane $x\-y$ for $M=1.936M_\odot$, $J=4.498$,
$b=-0.3080$, $E=0.95$ and $L=-9.0$ is presented. This
configuration is very similar to that presented in the first case
and is the commonest shape found. The geodesic behavior inside of
the bounded region for these potentials is presented in the
Fig.\ref{MS1}(b) and Fig.\ref{MS2}(b), respectively. From there
can be seen that the motion is completely regular.

In order to found chaotic behavior, we let ourself modify sightly
the values of the $b$ parameter changing the quadrupole deformation
and also the properties of the test particle throughout the change
in its energy and angular momentum values. Introducing this changes
we found potential like the presented in Fig.\ref{MS3}(a) and
Fig.\ref{MS4}(a).

In Fig.\ref{MS3}(a), we present the boundary contour for
$M=1.936M_\odot$, $J=4.498$, $b=0.8$, $E=0.96$ and $L=-9.9$
observing that there are two very small escape regions and two
disconnected bounded regions. In Fig.\ref{MS4}(a) boundary contour
for $\Phi=0$ in the plane $x\-y$ for $M=1.936M_\odot$, $J=4.498$,
$b=0.8$, $E=0.971$ and $L=-9.3$ is presented. This configuration
conserves the tiny scape regions but the confined regions are
connected. The geodesic behavior inside of the bounded region in
the phase space for these potentials is presented in the
Fig.\ref{MS3}(b) and Fig.\ref{MS4}(b), respectively. From there
one can notice the presence of a mixed phase space --chaotic and
regular-- to the left but only regular to the right in both cases.

\section{Concluding Remarks}
\label{sec:Conclusions}

We have constructed the Poincar\'e sections for a lot of possible
combinations of energy and angular momentum (and deformation
parameter in the Manko {\it {et al.}} case) which confine the motion
of the test particles orbiting around two types of sources.

We concluded that is apparently impossible to find chaotic
geodesics around a source described for the {\small \sc{TS}}
solution and that the stability of the geodesics does not depend
on any way of the relative spin direction of the center of
attraction nor of the angular momentum of the test particle. This
result shows numerical evidences of the existence of only
integrable geodesics for this system. In other words, the case of
the test particle turning around a Tomimatsu-Sato source type is
completely integrable. This work attempts to complete the study of
the geodesic dynamics in the well known trilogy of axially
symmetric solutions (Schwarzschild, Kerr, Tomimatsu Sato
$\delta=2$).

On the other hand, we use the Manko {\it {et al.}} solution with
the idea of analyze the geodesic dynamics of test particles in
rapidly rotating neutron stars with equation of state FPS. For
this we took the values presented by Berti and Stergioulas for the
parameters describing the multipolar structure of neutron stars.
In this case we only found regular geodesics like in the precedent
metric. In order to find chaotic behavior, we modify the values of
the deformation parameter. Introducing these changes we found
potentials with two bounded regions, in which the Poincar\'e
sections exhibit mixed phase spaces, i.e., phase spaces when some
of the KAM tori survives inside a chaotic sea.

The main result in the Letelier's work was the existence of
chaotic geodesics in the geometry that characterizes the prolate
case. In the oblate case these orbits appear to be regular. In our
case we found chaotic geodesics (in the Manko's solution) only in
the geometry that characterizes the oblate case. In
\cite{LetelPRE} Letelier {\it et al.} analyzed the influence of
the introduction of multipolar terms corresponding to the
quadrupole deformation in the Schwarzschild and Kerr solution
finding chaotic behavior for some values of the deformation.
Considering the former statement and based on the numerical
evidence of our work we could conjecture that the parameter of
arbitrary deformation is which what induces the ergodic motion for
uncharged test particles orbiting general relativistic vacuum
sources.

\section*{Acknowledgments}
J.D.S.-G. and L.A.P. acknowledge gratefully for financial support
from COLCIENCIAS, Colombia. This work also was supported by
Project No. 5116 (DIEF—Ciencias of the Universidad Industrial de
Santander, Colombia).


\suppressfloats

\begin{thebibliography}{20}
\bibitem{note1} Work along this line will be soon reported.
\end{thebibliography}
\end{document}